\documentclass[twocolumn,prl,aps]{revtex4}

\newcommand{\be}{\begin{equation}} \newcommand{\ee}{\end{equation}} 
\newcommand{\bea}{\begin{eqnarray}}\newcommand{\eea}{\end{eqnarray}}

\begin{document}
\preprint{quant-ph/yymmxxx}
\title{ Exactly Solvable Quasi-hermitian Transverse Ising Model}
\author{ Tetsuo Deguchi} \email{deguchi@phys.ocha.ac.jp}
\affiliation{ Department of Physics, Graduate School of Humanities and
Sciences,\\ Ochanomizu University,
2-1-1 Ohtsuka, Bunkyo-ku,\\ Tokyo 112-8610, Japan}
\author{ Pijush K. Ghosh} \email{pijushkanti.ghosh@visva-bharati.ac.in}
\affiliation{Department of Physics, Siksha-Bhavana,\\
Visva-Bharati University,\\
Santiniketan, PIN 731 235, India.}
\begin{abstract} 
A non-hermitian deformation of the one-dimensional transverse Ising model is 
shown to have the property of quasi-hermiticity. The 
transverse Ising chain is obtained from the starting non-hermitian 
Hamiltonian through a similarity transformation. 
Consequently, both the models have
identical eigen-spectra, although the eigen-functions are different. The
metric in the Hilbert space, which makes the non-hermitian model unitary
and ensures the completeness of states,
has been constructed explicitly. Although the longitudinal correlation
functions are identical for both the non-hermitian and the hermitian
Ising models, the difference shows up in the transverse correlation
functions, which have been calculated explicitly and are not always
real. A proper set of hermitian spin operators in the Hilbert space of
the non-hermitian Hamiltonian has been identified, in terms of which
all the correlation functions of the non-hermitian Hamiltonian become
real and identical to that of the standard transverse Ising model.
Comments on the quantum phase transitions in the non-hermitian model have
been made.
\end{abstract}
\maketitle

The discovery of a class of non-hermitian Hamiltonians admitting entirely
real spectra has generated a renewed interest in the study of 
quantum physics\cite{bend,ali,quasi,ddt,kumar,swan,multi,me,korff,song}.
The reality 
of the entire spectra is related to an underlying unbroken combined
Parity(${\cal{P}}$) and Time-reversal(${\cal{T}}$) symmetry\cite{bend}
and/or quasi-hermiticity\cite{ali,quasi} of the non-hermitian Hamiltonian.
Apart from a very few known examples\cite{swan,multi,me,korff},
one of the major technical
difficulties in the study of ${\cal{PT}}$ symmetric and/or
quasi-hermitian quantum physics is to find the appropriate
basis with respect to which the non-hermitian Hamiltonian becomes
hermitian. The description of a non-hermitian Hamiltonian admitting
entirely real spectra is incomplete in absence of such a basis,
since neither the unitarity nor the completeness of states are guaranteed.
It may be noted here that the completeness of states is an essential
criterion to claim a Hamiltonian to be exactly solvable.

The purpose of this letter is to introduce and study an exactly
solvable non-hermitian Hamiltonian of the type of transverse Ising 
model that admits entirely real spectra. In particular, we consider a
non-hermitian Hamiltonian and map it to the transverse Ising model
through a similarity transformation. Consequently, both the models
have identical spectra. We find the metric in the Hilbert space of
the non-hermitian Hamiltonian that is required to make the theory unitary
and to ensure the completeness of states.  We show that the $n$-point
longitudinal correlation function of the non-hermitian Hamiltonian is identical
to that of the standard hermitian transverse Ising model. We also calculate
the two-point transverse correlation functions of the non-hermitian
Hamiltonian exactly that reduces to that of the transverse Ising model
in the hermitian limit. However, the transverse correlation functions
are not always real. We identify a proper set of hermitian spin operators
in the Hilbert space of the non-hermitian Hamiltonian in terms of which
all the correlation functions of the non-hermitian Hamiltonian become
real and identical to that of the standard transverse Ising model.

There are many physical applications of the transverse Ising chain
such as quantum phase transitions and finite-temperature
crossovers \cite{sachdev,bkc,jacek}. The transverse Ising model has
also been studied\cite{qent} extensively from the viewpoint of quantum
entanglement and its connection with quantum phase transition.
We may thus expect that the non-hermitian transverse Ising Hamiltonian
gives an explicit and concrete example of a non-hermitian exactly solvable
many-body system and should be useful for studying some interesting 
properties of non-hermitian quantum systems explicitly. 

Non-hermitian quantum many-body systems are closely related to 
 several important topics in other subjects. For instance, 
non-hermitian quantum spin chains correspond to two-dimensional
classical systems with positive Boltzmann weights. In exactly
solvable models, the non-hermitian XY and XXZ spin chain Hamiltonians 
with Dzyaloshinsky-Moriya interactions commute with  the transfer 
matrix of the six-vertex model in the presence of an electric
field \cite{McCoy-Wu}, and the integrable chiral Potts model in the
most general case leads to a non-hermitian quantum Hamiltonian 
(see for review, \cite{chiralPotts,Albertini}).  Moreover,
non-hermitian asymmetric XXZ spin chains related to the one dimensional
diffusion models have been studied extensively in nonequilibrium 
statistical mechanics \cite{ASEP}. The inherent pseudo-hermiticity
of these spin models has been discovered very recently \cite{quasime}
which allows a unitary description with a modified inner product
in the Hilbert space.  Further, a non-hermitian quantum Ising spin
chain in one dimension \cite{vg} is known to be related to the celebrated
Yang-Lee model\cite{yl} that aptly describes ordinary second order phase
transitions. The non-hermiticity of the spin chain arises due
to the inclusion of an external complex magnetic field and an analysis based
on minimal conformal field theories is available\cite{vg,cardy}.
Very recently, pesudo-hermiticity of the non-hermitian Ising chain of 
Ref. \cite{vg} has been studied for a finite number of sites in
Ref. \cite{andreas} and any result for an arbitrary number of sites is still
lacking. The non-hermitian quantum Ising chain that is considered in this
paper is different from that of Ref. \cite{vg} and an exact description
for an arbitrary number of sites is possible.


Let us now 
consider the transverse Ising Hamiltonian in the following modified form,
\be
H = - \sum_{i=1}^N \left( J S_i^z S_{i+1}^z + \epsilon_1 S_i^+ +
\epsilon_2 S_i^- \right)
\ee
\noindent where $S_i^z, S_i^{\pm}=
%
%
 \left( S_i^x
\pm i S_i^y \right)$ are the spin-variables. 
The spin-variables can be represented in terms of the Pauli matrices
$\sigma^{\pm}= \frac{1}{2} \left ( \sigma^x \pm i \sigma^y \right ), \sigma^z$
and the $ 2 \times 2$ identity matrix $I$ as,
\bea
&& S_i^{z} = I \otimes \dots \otimes I \otimes \frac{1}{2}  
\sigma^z \otimes I
\otimes \dots \otimes I\nonumber \\
&& S_i^{\pm} = I \otimes \dots \otimes I \otimes \sigma^{\pm} \otimes I
\otimes \dots \otimes I,
\label{pauli}
\eea
\noindent where $\sigma^{\pm}$ and $\sigma^z$ are in the $i$-th position.
The parameter $J$ is real, however, $\epsilon_{1,2}$ are complex.
Thus, the Hamiltonian is non-hermitian for $\epsilon_1 \neq \epsilon_2^*$,
where a $^*$ denotes the complex conjugation. The Hamiltonian $H$ and
its adjoint $H^{\dagger}$ are related to each other through the transformation
$\epsilon_1 \leftrightarrow \epsilon_2^*$. The standard transverse Ising
model is recovered when both $\epsilon_1$ and $\epsilon_2$ are real
and $\epsilon_1= \epsilon_2$. Even for the hermitian case, i.e.
$\epsilon_1=\epsilon_2^*$, $H$ can be mapped to the standard transverse
Ising model through an unitary transformation. However, for general
$\epsilon_1$ and $\epsilon_2$, $H$ can not be mapped to a hermitian
Hamiltonian by using an unitary transformation.

The Hamiltonian $H$ can be mapped to a hermitian Hamiltonian through a
similarity transformation. To do so, let us introduce the operator $\rho$
and its inverse $\rho^{-1}$ in the following way,
\bea
&& \rho_i = \gamma^{-\frac{1}{2}} S_i^{+} S_i^{-} \ +
\ \gamma^{\frac{1}{2}} S_i^- S_i^+,\nonumber \\
&& \rho_i^{-1} = \gamma^{\frac{1}{2}} S_i^{+} S_i^{-} \ +
\ \gamma^{-\frac{1}{2}} S_i^- S_i^+,\nonumber \\
&& \rho= \prod_{i=1}^N \rho_i, \ \ \ \ 
\rho^{-1}= \prod_{i=1}^N \rho_i^{-1},
\eea
\noindent where $\gamma=\sqrt{\frac{\mid \epsilon_1
\mid}{\mid \epsilon_2 \mid}}$.
The ordering of $\rho_i$'s is not required in the definition of
the positive-definite operators $\rho$ and $\rho^{-1}$,
since $[\rho_i, \rho_j]=0$ for $ i \neq j$.  Using the following identities,
\bea
&& \rho \ S_i^{z} \ \rho^{-1} = \ S_i^{z},\nonumber \\
&& \rho \ S_i^{\pm} \ \rho^{-1} = \gamma^{\mp 1} \ S_i^{\pm},
\label{iden}
\eea
\noindent one can easily check that,
\bea
h & = & \rho H \rho^{-1} \nonumber \\
 & = & - \sum_{i=1}^N J S_i^z S_{i+1}^z\nonumber \\
& - & \beta \sum_{i=1}^N \left ( e^{i \arg(\epsilon_1)} \ S_i^+
+  e^{i \arg(\epsilon_2)}
\ S_i^- \right ).
\label{qh}
\eea
\noindent The parameter $\beta$ appearing in $h$ is defined as,
$\beta \equiv \sqrt{ {\mid \epsilon_1 \mid} {\mid \epsilon_2 \mid}}$.
The Hamiltonian $h$ is hermitian when the following condition holds true, 
\be
\arg (\epsilon_1) + \arg (\epsilon_2) = 2 k \pi, \ \ k=0, 1, 2, \dots
\label{eqality}
\ee
\noindent Thus, Eq. (\ref{eqality}) is the condition for $H$ to be
quasi-hermitian, i.e. related to the hermitian $h$ through the similarity
transformation.
A counter-clockwise rotation on the $S_x-S_y$-plane around the
$S_z$-axis by an angle $\xi \equiv \arg(\epsilon_1)=- \arg(\epsilon_2)$,
followed by a clock-wise rotation by an angle $\frac{\pi}{2}$ around $S_y$-axis
for each spin transform $h$ into the standard form of the transverse
Ising model. To this end, we introduce an operator $U$ as,
\be
U=\prod_{i=1}^N e^{i \frac{\pi}{2} S_i^y} \prod_{j=1}^N
e^{-i \xi S_j^z},
\label{uni}
\ee
\noindent which transforms $S_i^{x,y,z}$ in the following way:
\bea
U S_i^z U^{-1} & = & - S_i^x,\nonumber \\
U S_i^y U^{-1} & = & S_i^y \cos \xi - S_i^z \sin \xi,\nonumber \\
U S_i^x U^{-1} & = & S_i^z \cos \xi + S_i^y \sin \xi.
\label{uni1}
\eea
\noindent Using the above identities, $h$ can be transformed to $\cal{H}$,
\bea
{\cal{H}} & = & U h U^{-1}\nonumber \\
&=& - \sum_{i=1}^N \left ( J S_i^x S_{i+1}^x + \beta S_i^z \right )
\label{modi}
\eea
\noindent which is the standard form of the transverse Ising model.
The entire energy spectra of $H$ is real and identical to that of
${\cal{H}}$, since a similarity transformation can not change the eigenvalues. 
However, as we will see below, the difference between $H$ and the transverse
Ising Model (i.e.  $H$ with real $\epsilon_1=\epsilon_2$) shows up in the
eigenstates and the transverse correlation functions.

The transverse Ising model ${\cal{H}}$ is exactly solvable and the different
correlation functions can be calculated explicitly\cite{lsm,pefuty,mc,perk}.
Using the Jordan-Wigner transformation,
the Hamiltonian ${\cal{H}}$ can be transformed to a fermionic
Hamiltonian which is quadratic in the fermionic annihilation and creation
operators.
The resulting fermionic Hamiltonian can be further diagonalized in terms of
a new set of canonical Fermi-operators\cite{lsm}.
It is worth mentioning here that the direct
application of the Jordan-Wigner transformation to $H$ produces a fermionic
Hamiltonian with non-local, non-hermitian interaction. However, the
operators $\rho$ and its inverse $\rho^{-1}$, defined appropriately in terms
of fermionic annihilation and creation operators transform $H$ to the
fermionic version of ${\cal{H}}$ that is local and hermitian.

If $|\psi_n\rangle$ constitutes a complete set of orthonormal eigenstates
of the hermitian Hamiltonian ${\cal{H}}$ with energy eigenvalue $E_n$, 
then,
\be
| \phi_n \rangle = \left ( U \ \rho \right )^{-1} | \psi_n \rangle,
\ \ |\chi_n \rangle = \left ( \rho \ U^{-1} \right ) | \psi_n \rangle
\ee
\noindent are the eigenstates of $H$ and its adjoint $H^{\dagger}$,
respectively. It may be noted here that both $H$ and $H^{\dagger}$
share the same energy eigenvalue $E_n$ with ${\cal{H}}$. However,
neither $|\phi_n\rangle$ nor $|\chi_n\rangle$ constitute a complete set
of orthonormal basis vectors. Consequently, with the standard norm in the
Hilbert space,
the time-evolution of $H$(or $H^{\dagger}$) is not unitary, although the
entire eigen-spectra are real. As is evident from Eqs.
(\ref{qh}-\ref{modi}), $H$ is a quasi-hermitian operator. Thus, the
Hilbert space of $H$ admits a bi-orthogonal structure,
\be
\langle \chi_n|\phi_m \rangle = \delta_{nm}, \ \
\sum_n |\chi_n \rangle \langle \phi_n | = 1.
\ee 
\noindent The completeness of states can be accomplished if the
inner-product in the Hilbert space is modified as\cite{ali},
\be
\langle \langle u, v \rangle \rangle_{\eta_+} := \langle u, \eta_+ v \rangle,
\ \ \eta_+ := \rho^2.
\ee
\noindent With this new inner-product in the Hilbert space, the expectation
value of an operator $\hat{O}$ can be calculated as,
\be
\langle \langle \hat{O} \rangle \rangle_{\eta_+} \equiv
\langle \phi_n | \eta_+ \hat{O} | \phi_n \rangle = \langle \psi_n |
\left ( U \rho \right )  \hat{O} \left (U \rho \right ) ^{-1} |\psi_n \rangle.
\label{ip}
\ee
\noindent We will be using the above expression to calculate $n$-point
correlation function of $H$. The standard inner product $\langle u,
v \rangle$ will be used to calculate the correlation function of the
hermitian Hamiltonian ${\cal{H}}$.

An $n$-point($n \leq N$) longitudinal correlation function of $H$
for the $m$th eigenstate can be related to the correlation function
of the transverse Ising model in the following way:
\be
\langle \langle S_{i_1}^z S_{i_2}^z \dots S_{i_n}^z
\rangle \rangle_{\eta_+}
= (-1)^n \langle \psi_m | S_{i_1}^x S_{i_2}^x \dots S_{i_n}^x 
| \psi_m \rangle,
\ee
\noindent where any two of the indices $i_k$ are not equal. 
Identifying $S_i^z$ of $H$ with $-S_i^x$ of ${\cal{H}}$, we observe
that the longitudinal correlation functions for these two systems
are identical.
However, the $n$-point transverse correlation 
functions of $H$ and $\cal{H}$ differ from each other.
Let us introduce a complex parameter
$z \equiv \ln \gamma + i \xi, \gamma > 0$
in terms of $\gamma$ and $\xi$. We also
introduce two operators $Q_{i_1, i_2, \dots, i_n}$ and
$\tilde{Q}_{i_1, i_2, \dots i_n}$ as,
\bea
Q_{i_1 i_2 \dots i_n} & = & \prod_{j=1}^n \left ( \cosh z \, S_{i_j}^z -
i \sinh z \, S_{i_j}^y \right )\nonumber \\
\tilde{Q}_{i_1 i_2 \dots i_n} & = & \prod_{j=1}^n
\left ( i \sinh z \, S_{i_j}^z +
\cosh z \, S_{i_j}^y \right ). 
\eea
\noindent The transverse correlation functions of $H$
and ${\cal{H}}$ can now be related as,
\bea
\langle \langle S_{i_1}^x S_{i_2}^x \dots S_{i_n}^x
\rangle \rangle_{\eta_+} & = &
\langle \psi_m| Q_{i_1 i_2 \dots i_n} |\psi_m \rangle\nonumber \\
\langle \langle S_{i_1}^y S_{i_2}^y \dots S_{i_n}^y
\rangle \rangle_{\eta_+} & = &
\langle \psi_m| \tilde{Q}_{i_1 i_2 \dots i_n} |\psi_m \rangle.
\eea
\noindent In general, for $\gamma \neq 1$, the correlation functions are
complex. For example, the one-point correlation functions in
the ground-state $|\psi_0 \rangle$ can be be evaluated as,
\be
\langle \langle S_i^x \rangle \rangle_{\eta_+}  = 
\cosh z \, M_i^z; \ \
\langle \langle S_i^y \rangle \rangle_{\eta_+}  = 
i \sinh z \, M_i^z
\ee
\noindent where $M_i^{x,y,z} \equiv
\langle \psi_0 | S_i^{x,y,z} | \psi_0 \rangle$
and we have used the result\cite{mc} that $M_i^y=0$ 
for arbitrary $\lambda \equiv \frac{J}{\beta}$. It may be noted that
for $\gamma \neq 1$,
$\langle \langle S_i^x \rangle \rangle_{\eta_+}$ is real only for
$\xi = n \pi$, 
\be
\langle \langle S_i^x \rangle \rangle_{\eta_+} = 
(-1)^n \cosh(\ln \gamma) \, M_i^z,
\ee
\noindent while $\langle \langle S_i^y \rangle \rangle_{\eta_+}$ is real
only for $\xi = (2 n + 1 ) \frac{\pi}{2}$,
\be
\langle \langle S_i^y \rangle \rangle_{\eta_+}  = 
(-1)^{n+1} \cosh(\ln \gamma) \, M_i^z,
\ee
where $n$ is either zero or a positive integer. 
It is expected that for $H$ to describe a  physical theory,
at least both the magnetization along $X$ and $Y$ direction should be
real, which is not the case for a fixed $\xi$ and $\gamma \neq 1$.
This is certainly an unwanted feature. 

The two-point diagonal correlation functions have the following form,
\bea
\langle \langle S_i^x S_j^x \rangle \rangle_{\eta_+} & = &
\cosh^2 z \, C_{ij}^z
- \sinh^2 z \, C_{ij}^y\nonumber \\
\langle \langle S_i^y S_j^y \rangle \rangle_{\eta_+} & = &
- \sinh^2 z \, C_{ij}^z,
+ \cosh^2 z \, C_{ij}^y
\label{2p}
\eea
\noindent where
$C_{ij}^{x,y,z} \equiv \langle \psi_0| S_i^{x,y,z} S_j^{y,z}|
\psi_0 \rangle$ and we have used the result\cite{mc}
$\langle \psi_0 | S_i^z S_j^y | \psi_0 \rangle =
\langle \psi_0 | S_i^y S_j^z | \psi_0 \rangle=0$
for arbitrary $\lambda$. For $\gamma \neq 1$,
both $\langle \langle S_i^x S_j^x \rangle \rangle_{\eta_+}$ and
$\langle \langle S_i^y S_j^y \rangle \rangle_{\eta_+}$ are real for
$\xi = \frac{n \pi}{2}$.
In particular, for $\xi= n \pi$:
\bea
\langle \langle S_i^x S_j^x \rangle \rangle_{\eta_+} & = &
\cosh^2(\ln \gamma) \, C_{ij}^z - \sinh^2(\ln \gamma) \, C_{ij}^y\nonumber \\
\langle \langle S_i^y S_j^y \rangle \rangle_{\eta_+} & = &
- \sinh^2(\ln \gamma) \, C_{ij}^z + \cosh^2(\ln \gamma) \, C_{ij}^y 
\nonumber \\ 
\eea
\noindent and for $\xi=(2n+1) \frac{\pi}{2}$:
\bea
\langle \langle S_i^x S_j^x \rangle \rangle_{\eta_+} & = &
- \sinh^2(\ln \gamma) \, C_{ij}^z + \cosh^2(\ln \gamma) \, C_{ij}^y 
\nonumber \\
\langle \langle S_i^y S_j^y \rangle \rangle_{\eta_+} & = &
\cosh^2(\ln \gamma) C_{ij}^z - \sinh^2(\ln \gamma) \, C_{ij}^y.
\eea
\noindent The diagonal correlation functions explicitly depend
on $\gamma$ and reproduce the known results in the hermitian limit
$\gamma=1$.

The off-diagonal two-point correlation functions have the following form,
\bea
\langle \langle S_i^x S_j^y \rangle \rangle_{\eta_+} & = &
{\frac i 2} \sinh 2z \left ( C_{ij}^z - C_{ij}^y \right )\nonumber \\
\langle \langle S_i^x S_j^z \rangle \rangle_{\eta_+} & = &
0 
\nonumber \\
\langle \langle S_i^y S_j^z \rangle \rangle_{\eta_+} & = &
0
\eea
\noindent where we have used the result \cite{mc} $\langle \psi_0|
S_i^x S_j^y | \psi_0 \rangle =0 $ and 
$\langle \psi_0 | S_i^z S_j^x | \psi_0 \rangle =0$ 
for arbitrary $\lambda$. It may be
noted that $\langle \langle S_i^x S_j^y \rangle \rangle_{\eta_+}$ is
complex for $\gamma \neq 1$ and arbitrary $\xi$. 
Both
$\langle \langle S_i^x S_i^z \rangle \rangle_{\eta_+}$ and
$\langle \langle S_i^y S_i^z \rangle \rangle_{\eta_+}$ vanishes
for arbitrary $\lambda$. 
%
\noindent Other $n$-point correlation functions
with higher values of $n$ may be calculated in the same way. Some of
them may become complex 
for $\gamma \neq 1$. 

One of the major criticisms of the above results could be
that all the one- and two-point correlation functions are not real
simultaneously for a fixed $\xi$ and $\gamma \neq 1$. The reason could
be traced to the fact that although $S_i^z$ are hermitian in the Hilbert
space of $H$ that is endowed with the metric $\eta_+$, the same is not
true for the spin variables $S_i^x$ and $S_i^y$. As a result, in general,
different correlation functions involving $S_i^x$ and $S_i^y$ are complex.

It is worth mentioning here that a common problem in the study
of pseudo-hermitian and ${\cal{PT}}$ symmetric quantum physics is that
although the entire energy eigen values of a non-hermitian Hamiltonian
may become real with unitary time evolution, expectation values of other
physical quantities of interest may not be real. Thus, a complete description
of non-hermitian Hamiltonian is not imminent. A common understanding
in this regard is that the metric $\eta_+$ is not unique and a more
general metric in the Hilbert space of $H$ may be found so that all the
correlation functions are real along with the eigenvalues. Since, a 
generalized metric that gives a complete description of $H$ is not guaranteed
a priory, identification of a proper set of operators those are hermitian
with respect to $\eta_+$ may give rise to a complete description of $H$.
General prescription in this regard is already known\cite{ali}.
To this end, we introduce a new set of spin-operators,
\bea
T_i^x & := & - S_i^z,\nonumber \\
T_i^y & := & - i S_i^x \sinh z + S_i^y \cosh z\nonumber \\
T_i^z & := & S_i^x \cosh z + i S_i^y \sinh z,
\eea
\noindent which satisfy the standard $SU(2)$ algebra.
It should be noted here that $T_i^{x,y}$ are not hermitian
in the sense of Dirac-hermiticity, i.e $\langle u, T_i^{x,y} v \rangle
\neq \langle T_i^{x,y} u, v \rangle$. However, the operators
$T_i^{x,y,z}$ are hermitian in the Hilbert space of $H$ with respect to
the metric $\eta_+$. The Hamiltonian $H$ can be rewritten as,
\be
H = - \sum_{i=1}^N \left( J T_i^x T_{i+1}^x + \beta T_i^z \right).
\label{herm}
\ee
\noindent The Hamiltonian $H$ is hermitian, i.e.
$\langle u, \eta_+ H v \rangle = \langle H u, \eta v \rangle$.
Using the identities,
\be
\left ( U \rho \right ) T_i^{x,y,z} \left ( U \rho \right )^{-1}
 = S_i^{x,y,z},
 \ee
 \noindent and Eq. (\ref{ip}), it is easy to see that the n-point
 correlation functions of $H$ and ${\cal{H}}$ are now identical,
 \be
 \langle \langle T_{i_1}^p T_{i_2}^q \dots T_{i_n}^r \rangle \rangle_{\eta_+}
 = \langle \psi_m | S_{i_1}^p S_{i_2}^q \dots S_{i_3}^r |\psi_m \rangle,
 \ee
 \noindent where the superscripts $p, q, r$ can be identified with $x, y, z$.
This implies that the Hamiltonian $H$ that is non-hermitian with respect to
the condition of Dirac-hermiticity has in fact a consistent and complete
description in terms of the new spin-operators $T_i^{x,y,z}$ those are
hermitian in the Hilbert space of $H$ that is endowed with the metric $\eta_+$.
Moreover, energy eigenvalues and different correlation functions of $H$
and ${\cal{H}}$ are identical.

One pertinent question that seems unavoidable at this point is whether the
particular choice of the positive-definite metric $\eta_+$ has any role in
establishing identical $n$-point correlation functions for $H$ and ${\cal{H}}$.
It seems that the answer is in the negative as long as the proper
identification of the new set of spin operators is made for a given
positive-definite metric. In particular, for a given positive-definite metric
$\eta_+:= \Gamma^2$, the hermitian spin operators $\Sigma_i^{x,y,z}$ should
be chosen as,
\be
\Sigma_i^{x,y,z} := \Gamma^{-1} S_i^{x,y,z} \Gamma, \ \forall \ i,
\ee
\noindent which would automatically imply identical correlation functions
for $H$ and ${\cal{H}}$. This observation is important, since it gives a metric
independent description.

Finally, a few comments are in order:\\
(i) The transverse-field Ising model is known to posses a global phase flip
symmetry. The same symmetry is present in the pseudo-hermitian Hamiltonian $H$
also with the phase flip operator $K$ given by,
\be
K := \prod_{i=1}^N T_i^z.
\ee
\noindent The operator $K$ acts on $S_i^{x,y,z}$ in the following way:
\be
K S_i^z K^{-1}  =  - S_i^z,\ \
K S_i^{\pm} K^{-1} = e^{\mp 2 z} S_i^{\mp}.
\ee
\noindent The phase flip symmetry acts quite non-trivially on $S_i^{x,y}$.

The Krammers-Wannier duality of the standard transverse-field Ising model can
also be established for $H$. Defining a set of spin-operators which obey
$SU(2)$ algebra and are hermitian with respect to the modified inner product
in the Hilbert space,
\be
\tau_i^x := \prod_{k \leq i} T_k^z, \ \
\tau_i^y  := - T_i^y T_{i+1}^x \prod_{k \leq i-1} T_k^z, \ \
\tau_i^z := T_i^x T_{i+1}^x,
\ee
\noindent the Hamiltonian $H$ can be rewritten as,
\be
H = - \sum_{i=1}^N \left ( J \tau_i^z + \beta \tau_i^x \tau_{i+1}^x \right ).
\ee
\noindent Based on the standard arguments, $\lambda=1$ is determined as the
critical point/line.

(ii) The transverse Ising model undergoes quantum phase transition. Since
${\cal{H}}$ is related to $H$ through the similarity transformation, $H$
also undergoes quantum phase transition. The quantum critical line of
$H$ is determined by $\lambda = 1$ and it contains the quantum critical
point of ${\cal{H}}$. 

Near the critical line/point of the quantum phase transition, 
the equal-time correlations of the order parameter does not change 
for $H$ and ${\cal{H}}$. Here we recall that the longitudinal correlation 
functions are identical for both the non-hermitian and the hermitian
Ising models. However, the degree of quantum coherence among spins
$S_i^{x,y,z}$ should be different for $H$ and ${\cal{H}}$ due to the
difference in the transverse correlation functions. The description
of $H$ in terms of the spin operators $S_i^{x,y,z}$ is incomplete and improper,
since the correlation functions are not always real. A consistent and
complete description in terms of the spin operators $T_i^{x,y,z}$
is possible and both transverse and longitudinal correlations functions
are identical for $H$ and ${\cal{H}}$. Consequently, the order parameter
and the degree of quantum coherence near the critical point/line remains
the same for both the Hamiltonian.

(iii) The Hamiltonian $H$ is quasi-hermitian. A concept related to
quasi-hermiticity is pseudo-hermiticity. One can show that $H$ is
pseudo-hermitian,i.e., $H^{\dagger} = \theta H \theta^{-1}$, where
$\theta := \rho^2$. The operator $\theta$ and its inverse are evaluated
as given below,
\bea
&& \theta = \prod_{i=1}^N \theta_i =
\prod_{i=1}^N \left ( \gamma^{-1} S_i^{+} S_i^{-} \ +
\ \gamma S_i^- S_i^+ \right ),\nonumber \\
&& \theta^{-1} = \prod_{i=1}^N \theta_i =
\prod_{i=1}^N \left ( \gamma S_i^{+} S_i^{-} \ +
\ \gamma^{-1} S_i^- S_i^+ \right ).
\eea
\noindent Note that $[ \theta_i, \theta_j ] =0$ for $i \neq j$. Hence, 
the ordering of $\theta_i$'s are not required in the expression above.

(iv) There are many physically motivated generalizations of the transverse
Ising model\cite{bkc,jacek} and are important in the study of phase
transitions. A non-hermitian deformation of such models can be shown to be
quasi-hermitian. For example, consider the non-hermitian Hamiltonian,
\bea
H_1 & = & - \sum_{i=1}^N \left ( J_i S_i^z S_{i+1}^z +
K_i S_i^z S_{i+k}^z \right )\nonumber \\
& - & \sum_{i=1}^N \left ( \epsilon_{1,i} e^{i \xi_i} S_i^+ +
\epsilon_{2,i} e^{- i \xi_i} S_i^- \right ),
\eea
\noindent where $J_i, K_i, \xi_i, \epsilon_{1,i}, \epsilon_{2,i}$ are real
and $k$ is an integer satisfying $1 < k < N$. 
The Hamiltonian $H_1$ is hermitian for
$\epsilon_{1,i} = \epsilon_{2,i}, \forall \ i$. Define the similarity
operator $\rho_1$ and its inverse as,
\bea
&& \rho_1 = \prod_{i=1}^N \left ( \gamma_i^{-\frac{1}{2}} S_i^{+} S_i^{-} \
+ \ \gamma_i^{\frac{1}{2}} S_i^- S_i^+ \right ),\nonumber \\
&& \rho_1^{-1} = \prod_{i=1}^N \left ( \gamma_i^{\frac{1}{2}} S_i^{+} S_i^{-} \
+ \ \gamma_i^{-\frac{1}{2}} S_i^- S_i^+ \right ),
\eea
\noindent where $\gamma_i \equiv \sqrt{\mid
\frac{\epsilon_{1,i}}{\epsilon_{2,i}} \mid}$. The non-hermitian $H_1$ can
be mapped to a hermitian ${\cal{H}}_1$ through the similarity
transformation,
\bea
{\cal{H}}_1 & = & \rho_1 H_1 \rho_1^{-1}\nonumber \\
& = & - \sum_{i=1}^N \left ( J_i S_i^z S_{i+1}^z +
K_i S_i^z S_{i+k}^z \right ) \nonumber \\
& - & \sum_{i=1}^N \left [ \beta_i \left (
e^{i \xi_i} S_i^+ + e^{- i \xi_i} S_i^- \right ) \right ],
\eea
\noindent where $\beta_i \equiv \sqrt{{\mid \epsilon_{1,i} \epsilon_{2,i}
\mid}}$.Thus, $H$ is quasi-hermitian.

We have considered a non-hermitian deformation of the transverse Ising
model that is also quasi-hermitian. The transverse Ising model has been
obtained from the starting non-hermitian Hamiltonian through a similarity
transformation. Consequently, both the models have identical eigen-spectra,
although the eigen-functions are different. The metric in the Hilbert space,
which makes the non-hermitian model unitary and ensures a complete set of
states, has been constructed explicitly. Although the longitudinal correlation
functions are identical for both the non-hermitian and the hermitian
Ising models, the difference shows up in the transverse correlation
functions, which have been calculated explicitly. However, the transverse
correlation functions are not always real. In order to give a complete
and consistent description, we have identified a proper set
of hermitian spin operators in the Hilbert space of the non-hermitian
Hamiltonian in terms of which all the correlation functions of the
non-hermitian Hamiltonian become real and identical to that of the standard
transverse Ising model. The non-hermitian
Hamiltonian undergoes quantum phase transitions and it is expected
that around the quantum critical line both the order parameter and the
degree of quantum coherence of spins should be identical to that of
the standard transverse Ising model.

\acknowledgements{ We would like to thank J. H. H. Perk for
communications. PKG would like to thank Ochanomizu University for warm
hospitality during his visit under the JSPS Invitation Fellowship for
Research in Japan (S08042), where a part of this work has been carried out.
This work is partially supported by 
Grant-in-Aid for Scientific Research (C) No. 20540365.   
}

\end{document}